\documentclass[useAMS,usenatbib,letterpaper]{mn2e} 
\usepackage[totalwidth=480pt,totalheight=680pt,layoutvoffset=0.5cm]{geometry}
\usepackage[latin9]{inputenc}
\usepackage{color}
\usepackage{float}
\usepackage{url}
\usepackage{amsmath}
\usepackage{amssymb}
\usepackage{astrobib_mnras2e}
\usepackage{graphicx}
\usepackage{times} 
\usepackage{fixltx2e}

\makeatletter

\newcommand{\mvir}{M_{\rm vir}}
\newcommand{\mcoll}{M_{\rm coll}}
\newcommand{\rvir}{R_{\rm vir}}
\newcommand{\vmax}{V_{\rm max}}
\newcommand{\meanvmax}{\overline{V}_{\rm max}}

\newcommand{\ximm}{\xi_{\rm mm}}
\newcommand{\xihm}{\xi_{\rm hm}}
\newcommand{\bh}{b_{\rm h}}
\newcommand{\bhlin}{b^{\rm lin}_{\rm h}}

\newcommand{\msun}{M_{\odot}/h}
\newcommand{\mpch}{{\rm Mpc}/h}
\newcommand{\kpch}{{\rm kpc}/h}

\makeatother

\begin{document}

\title[The Scale-Dependence of Halo Assembly Bias]{The Scale-Dependence of Halo Assembly Bias}
\author[Sunayama et al.]
{Tomomi Sunayama$^{1,2}$, Andrew P. Hearin$^2$, Nikhil Padmanabhan$^{1,2}$, Alexie Leauthaud$^3$\\
$^1$Department of Physics, Yale University, P.O. Box 208120, New Haven, CT \\
$^2$Yale Center for Astronomy \& Astrophysics, Yale University, New Haven, CT\\
$^3$ Kavli IPMU (WPI), UTIAS, The University of Tokyo, Kashiwa, Chiba 277-8583, Japan
}

\maketitle
\begin{abstract}
The two-point clustering of dark matter halos is influenced by halo properties besides mass, a phenomenon referred to as {\em halo assembly bias}. Using the depth of the gravitational potential well, $\vmax,$ as our secondary halo property, in this paper we present the first study of the scale-dependence assembly bias. In the large-scale linear regime, $r\gtrsim10\mpch,$ our findings are in keeping with previous results. In particular, at the low-mass end ($\mvir<\mcoll\approx10^{12.5}\msun$), halos with high-$\vmax$ show stronger large-scale clustering relative to halos with low-$\vmax$ of the same mass; this trend weakens and reverses for $\mvir\gtrsim\mcoll.$ In the nonlinear regime, assembly bias in low-mass halos exhibits a pronounced scale-dependent ``bump''  at $500\kpch-5\mpch,$ a new result. This feature weakens and eventually vanishes for halos of higher mass. We show that this scale-dependent signature can primarily be attributed to a special subpopulation of {\em ejected halos}, defined as present-day host halos that were previously members of a higher-mass halo at some point in their past history. A corollary of our results is that galaxy clustering on scales of $r\sim1-2\mpch$ can be impacted by up to $\sim15\%$ by the choice of the halo property used in the halo model, even for stellar mass-limited samples. 
\end{abstract}

\section{Introduction}

The halo model  
provides a connection between dark
matter halos and galaxies, and it has been remarkably successful in
describing observations of galaxy clustering \citep[][for a recent review]{seljak00,vdboschBOOK}. In particular, the
Halo Occupation Distribution (HOD)\citep{berlind02,berlind03} and the Conditional Luminosity
Function (CLF)\citep{yang03} are the two most widely used models of the galaxy-halo
connection. These models start from the assumption that halo mass completely determines
the galaxy occupation statistics. In order to populate halos with
galaxies, the HOD specifies the probability $P(N|M)$ that a halo with
mass $M$ hosts $N$ galaxies, while the CLF models the mean
abundance $\Phi(L|M)$ of galaxies with luminosity $L$ in halos of  mass $M$.
These two models are interchangeable; integrating the CLF over luminosity
yields an HOD. Both models have been applied
extensively to observations in order to study the galaxy-halo connection
\citep{magliocchetti03,zehavi05a,yang_etal05,zheng07,vdBosch07,zheng09,skibba09,simon_etal09,ross10,zehavi11,leauthaud11b,leauthaud12,geach12,parejko13}
as well as cosmology \citep{tinker05,more_etal13,cacciato_etal13,mandelbaum13}.

However, the clustering of halos also exhibits a
dependence on additional properties beyond their mass 
\citep{gao_etal05,wechsler06,gao_white07,li_etal08,hahn_etal09}, a
phenomenon generically referred to as {\em halo assembly bias}. This can be 
traced back to the fact that halos of
the same mass in different environments have different assembly histories
and cluster differently. Having different assembly histories also affects
the internal structure of halos\citep{bullock01,wechsler02,hahn_etal07b,faltenbacher_white10}. This, in turn, results in a clustering dependence
on the structural properties of a halo, including the depth of its 
gravitational potential well, characterized by its maximum circular
velocity $V_{\rm max}$. The present work revisits this manifestation 
of halo assembly bias, and extends it down to smaller scales ($< 10 h^{-1} {\rm Mpc}$) 
than has been previously explored.

An alternative approach to connecting halos and galaxies is abundance
matching
\citep{kravtsov04a,vale_ostriker04,tasitsiomi_etal04,vale_ostriker06,conroy_wechsler09,
guo10,simha10,neistein11a,watson_etal12b,rod_puebla12,hearin_etal12b,kravtsov13,saito15}. 
In its simplest form, abundance matching posits a monotonic
relationship between a property of a galaxy (luminosity, stellar mass) 
and that of a halo (mass, potential well depth). By construction, 
such a relationship preserves the rank ordering of the galaxies and halos. 
The choice of the observationally-relevant halo property is a priori unknown; 
this uncertainty that can lead to significant systematic errors when using halo models 
to interpret galaxy clustering measurements \citep{zentner_etal13}. 

A recent parallel research effort has been revisiting the standard
subdivision of dark matter halos into host/sub-halos, a classification that naturally depends
on how the halo boundary is chosen. While the virial radius is the
most commonly chosen definition, recent work has demonstrated that the
environmental effects of halos extends well beyond the virial radius
\citep{wetzel_etal13, diemer14,adhikari14,wetzel_nagai14,more15}. 
In particular, \cite{wetzel_etal13} argue that these environmental effects
are due to ejected subhalos which orbit beyond the virial radius
of their hosts, and therefore get temporarily reclassified as host
halos. These studies 
argue that a more physically motivated boundary is the 
``splashback radius'' corresponding to the caustic from material 
just reaching its first apocentric passage. One of the chief results of the present work is 
that the halo assembly bias on small scales predominantly 
arises from this mis-classification.

The outline of this paper is as follows. Sec.~2 summarizes the simulations
we use in this work. Sec.~3 presents our primary results - characterizing
the dependence of the clustering of halos on $V_{\rm max}$; Sec.~4 explores
some of the implications of these results. We conclude in Sec.~5.

\section{Simulations}

We use the Bolshoi \citep{bolshoi_11} and MultiDark simulations \citep{riebe_etal11, prada_etal11}
\footnote{\url{http://www.MultiDark.org}%
} in this work; the combination of these simulations allows us to span
a large range in halo mass. We summarize key properties of these simulations
here. Both simulations were run with the Adaptive Refinement Tree
Code \citep{kravtsov_etal97,gottloeber_klypin08} assuming a flat
$\Lambda{\rm CDM}$ model with density parameters $\Omega_{m}=0.27$,
$\Omega_{\Lambda}=0.73$, $\Omega_{b}=0.0469$, and $\sigma_{8}=0.82$,
$n=0.95$, $h=0.70$. The Bolshoi simulation used $2048^{3}$ particles
in a $250h^{-1}{\rm Mpc}$ box with a force resolution of $1h^{-1}{\rm kpc}$,
giving a particle mass of $1.35\times10^{8}h^{-1}{\rm M_{\odot}}$,
while the MultiDark simulation used $2048^{3}$ particles in a $1h^{-1}{\rm Gpc}$
box with a force resolution of $7h^{-1}{\rm kpc}$ giving a particle
mass of $8.721\times10^{9}h^{-1}{\rm M_{\odot}}$. Dark matter halos
and subhalos are identified using the \texttt{ROCKSTAR} phase-space,
temporal halo finder \citep{rockstar} and merger trees are constructed
using the \texttt{CONSISTENT TREES} \citep{rockstar_trees} procedures%
\footnote{The catalogues used in this paper are publicly available at \url{http://hipacc.ucsc.edu/Bolshoi/MergerTrees.html}%
}. All of the results we consider here are at $z=0$.

In what follows, we use the virial masses and maximum circular velocities
(tagged ``\texttt{mvir}'' and ``\texttt{vmax}'') directly from
the halo catalogues. The halo mass and velocity functions start to
show incompleteness at $10^{10.4}h^{-1}M_{\odot}$ ($\sim200$ particles)
for the Bolshoi simulation, and $10^{12}h^{-1}M_{\odot}$ ($\sim100$
particles) for the MultiDark simulation. Below those masses, the halo mass distributions show unphysical drop-offs indicating incomplete mass resolution.

In addition to the standard classification of halos into host halos
(not within the virial radius of a more massive halo) and subhalos
(within the virial radius of a more massive halo), we further classify
host halos into ejected and non-ejected halos. Ejected halos, also
referred to as ``backsplash'' halos, are host halos whose main progenitor
was classified, at some point in its merger tree history, as a subhalo.
As we discuss below, these ejected halos have very different clustering
properties compared to their non-ejected counterparts. The ejected
fraction for the Bolshoi simulation is $\sim15.8\%$ at $10^{11}h^{-1}M_{\odot}$,
and drops to $\sim6.3\%$ at $10^{13}h^{-1}M_{\odot}$. The lower
mass resolution of the MultiDark simulation prevents us from making
this additional sub-classification. Accordingly, results that rely
on this split are restricted to the Bolshoi simulation and mass range.

\section{The Maximum Circular Velocity Dependence of Halo Clustering}

In this section we present our primary results. We begin in \S\ref{subsec:samples}
by describing the sample of halos we use throughout the paper, as
well as our method for how we categorize halos as having above- or
below-average circular velocities for their mass. In \S\ref{subsec:halobias},
we show the dependence of the clustering of halos on $V_{\rm max}$.

\subsection{Halo Sample Definitions}

\label{subsec:samples}

If the internal structure of a dark matter halo of mass $\mvir$ is
described by an NFW profile \citep{nfw97} of concentration $c,$
then its maximum circular velocity $\vmax$ is given by: 
\begin{equation}
\overline{V}_{{\rm max}}=0.465M_{{\rm vir}}^{1/3}\sqrt{G(\frac{4}{3}\pi\Delta_{{\rm h}}\rho_{{\rm crit}}\Omega_{m})^{1/3}\frac{c}{{\rm ln(1+c)-c/(1+c)}}}.\label{eq:vmax-mvir}
\end{equation}
As shown in \citep{klypin10a}, the median concentration-mass relation
$\bar{c}(\mvir)$ for $z=0$ Bolshoi halos is well-described by: 
\begin{equation}
{\rm log}_{10}\bar{c}=-0.097{\rm log_{10}M_{vir}}+2.148.\label{eq:concen}
\end{equation}
For every halo in the Bolshoi and MultiDark catalogs, we use its tabulated
$\mvir$ to compute $\bar{c}(\mvir),$ and then use the values $\bar{c}$
together with Eq.~\ref{eq:vmax-mvir} to compute $\overline{V}_{{\rm max}}$
for every halo. We will henceforth refer to halos with $\vmax<\meanvmax(\mvir)$
as ``low-$\vmax$ halos'', and halos with $\vmax>\meanvmax(\mvir)$
as ``high-$\vmax$ halos''. Thus a halo's high- or low-$\vmax$
designation refers to whether its true $\vmax$ value in the simulation
is above- or below-average for its mass.

\begin{figure}
\begin{center}
\includegraphics[width=0.9\columnwidth]{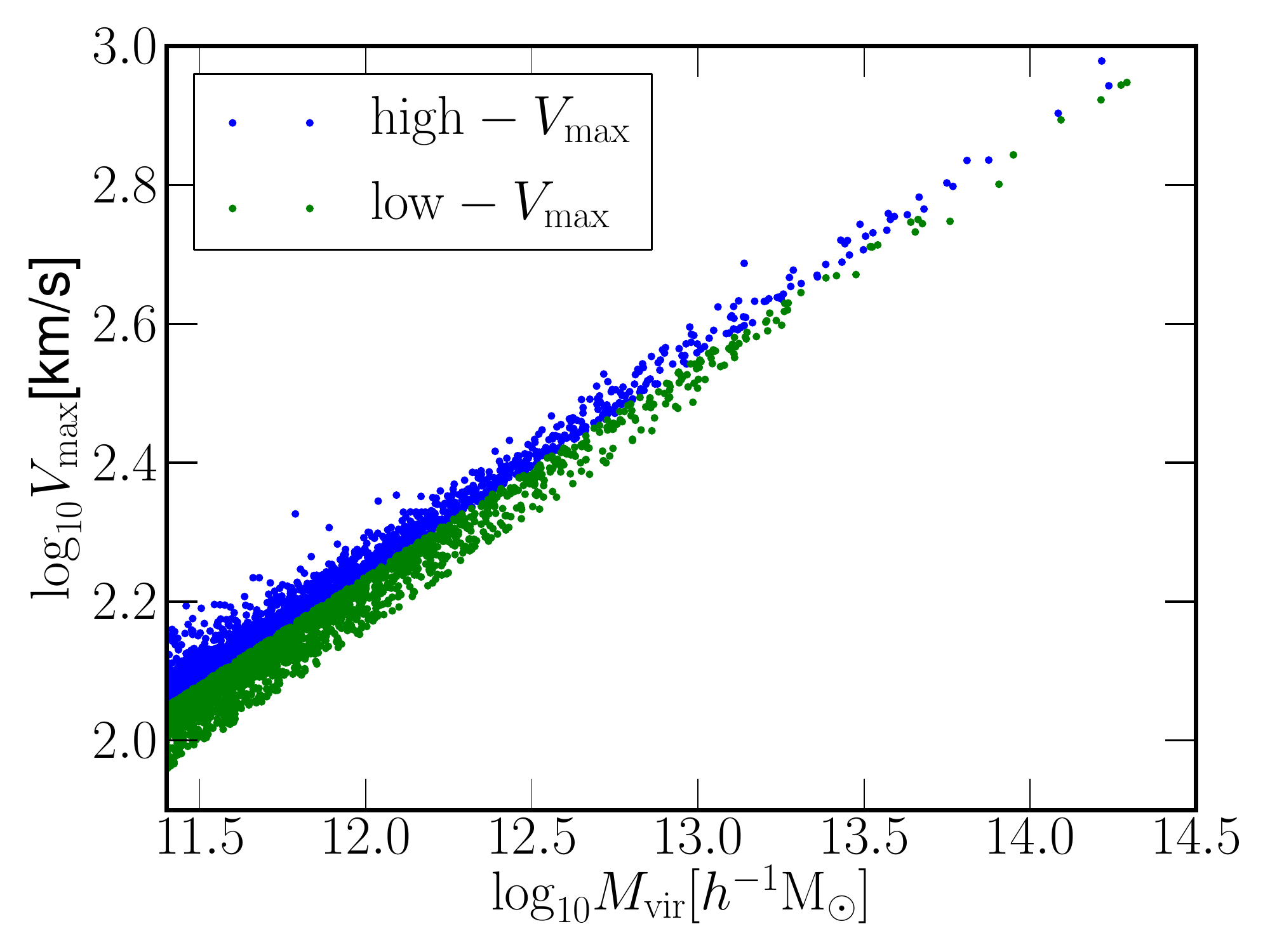}
\end{center}
\caption{\label{fig:vmax-mvir}Distribution of halo mass and maximum circular
velocity at $z=0.0$ for halos\textcolor{black}{{} from the Bolshoi
simulation.} The blue dots represent halos whose observed maximum
circular velocity, $V_{{\rm max,obs}}$, is greater than $\overline{V}_{{\rm max}}$,
while the green dots are the ones with smaller $V_{{\rm max,obs}}$
than $\overline{V}_{{\rm max}}$. The boundary between blue and green
dots correspond to $\overline{V}_{{\rm max}}$ computed from Eq. \ref{eq:vmax-mvir}.}
\end{figure}

Fig. \ref{fig:vmax-mvir} illustrates the distribution of Bolshoi
halos as a function of $\mvir$ and $\vmax.$ High-$\vmax$ halos
are shown in blue, low-$\vmax$ halos in green. The dividing line
between the two samples is defined by Eq.~\ref{eq:vmax-mvir}. We
find that our analytical approximation for $\meanvmax(\mvir)$ gives
a good description of the true median: for any fixed value of $\mvir,$
the high-$\vmax$ and low-$\vmax$ subsamples have very similar numbers
of objects.

Note that Fig. \ref{fig:vmax-mvir} shows a sharp lower bound on the
value of $\vmax$ at a given $\mvir,$ but no sharp upper bound. This
is ultimately due to the halo mass definition. The circular velocity
$V_{{\rm cir}}$ at the virial radius of any halo is $V_{{\rm cir}}(\rvir)\equiv V_{{\rm vir}}=\sqrt{G\mvir/\rvir}.$
Since the value of $\vmax$ tabulated in the halo catalog is computed
as the maximum value of $V_{{\rm cir}}$ over the entire profile of
the halo, formally $\vmax$ cannot exceed $V_{{\rm vir}}.$ This manifests
as the sharp lower bound seen in Fig. \ref{fig:vmax-mvir}.

\subsection{Halo Bias}

\label{subsec:halobias}

In this section we present our primary results for the clustering
properties of halos as a function of $\mvir$ and $\vmax.$ Clustering
strength is quantified by the two-point correlation function, $\xi(r).$
In all that follows, we will use $\ximm(r)$ to denote the auto-correlation
of the dark matter density field with itself, and $\xihm(r)$ to denote
the cross-correlation between a sample of halos and the underlying
density field.

Halos are biased tracers of the dark matter density field. We denote
this bias as $\bh,$ which is in general a function of spatial separation.
We define halo bias as 
\[
\bh(r)\equiv\xihm(r)/\ximm(r).
\]
On sufficiently large scales halo bias is approximately linear, and
$\bh(r)$ approaches a constant value $\bhlin.$

In order to measure the bias of a sample of simulated halos for both
Bolshoi and MultiDark simulations, we estimate $\ximm(r)$ and $\xihm(r)$ using
a random down-sampling of $10^{6}$ dark matter particles. For a given
sample of halos, we estimate the value of $\bhlin$ exhibited by the
sample as follows: 
\begin{equation}
b_{{\rm lin}}=\frac{1}{N_{{\rm bin}}}\sum_{i}(\xi_{\rm hm}(r_{i})/\xi_{\rm mm}(r_{i})).\label{eq:blindef}
\end{equation}
In Eq.~\ref{eq:blindef}, the sum is performed over $N_{{\rm bin}}=20$
separation bins $r_{{\rm i}}$ linearly spaced from $10h^{-1}{\rm Mpc}$
to $20h^{-1}{\rm Mpc}.$

In order to study the mass-dependence of halo bias, we bin our halos
into a sequence of $\mvir$ bins chosen such that there are the same
numbers of halos in each bin. For MultiDark, we select $2\times10^{5}$
halos for each bin; for Bolshoi we use $25000$ halos per bin. The
halos in each mass bin are categorized as high-$\vmax$ or low-$\vmax$
according to the method described in \S\ref{subsec:samples}. We
start using halos from the MultiDark simulation for $M_{{\rm vir}}>10^{12.2}h^{-1}{\rm M_{\odot}}$. 

\begin{figure}
\includegraphics[width=0.9\columnwidth]{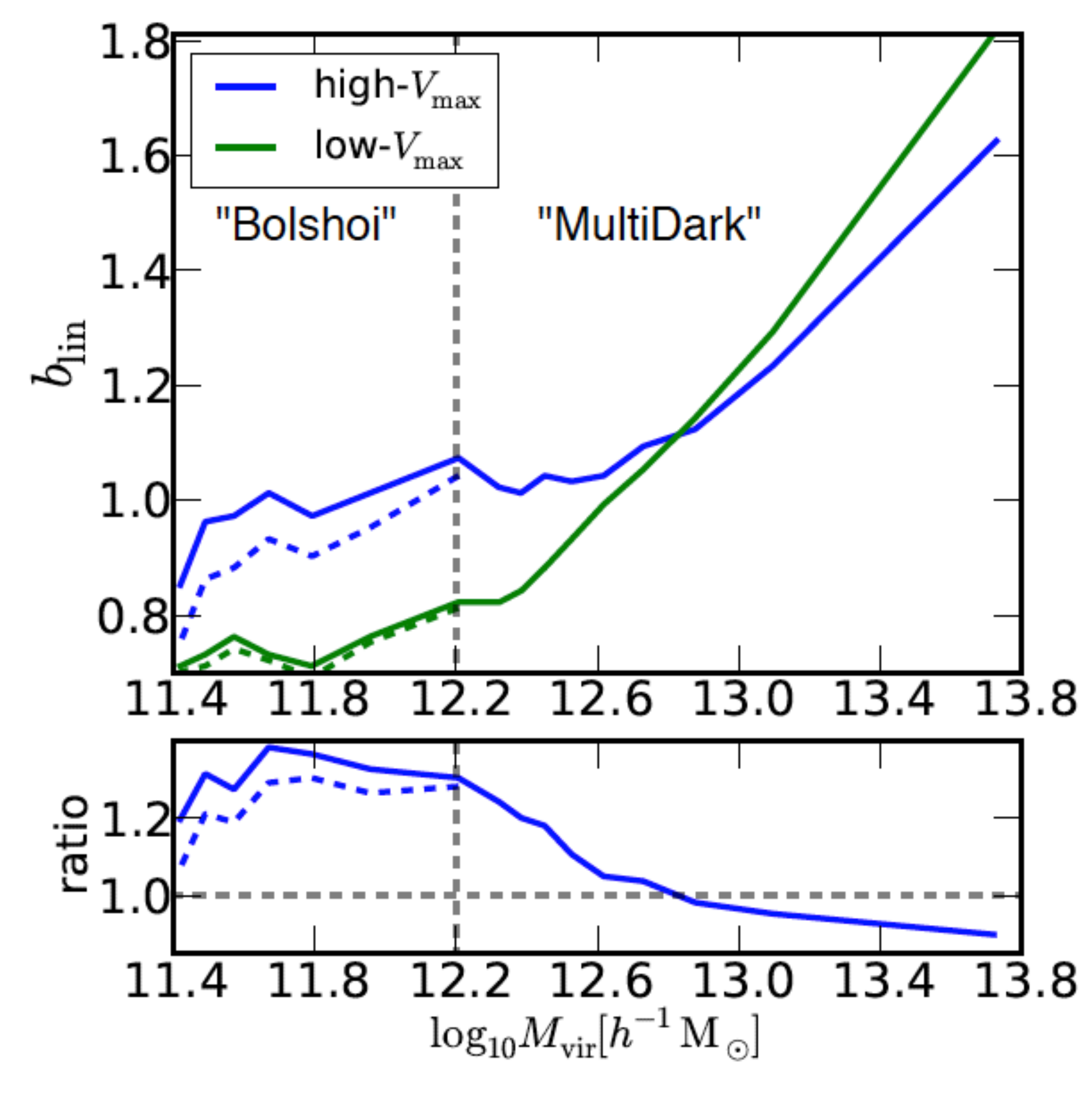}
\caption{\label{fig:linear-bias}Upper panel: Linear bias at $z=0$ as a
function of halo mass from the Bolshoi and the MultiDark simulations.
The vertical dashed line indicates the upper/lower limit of halo mass
for the Bolshoi/MultiDark simulations respectively. The blue lines
correspond to a linear bias for high-$V_{{\rm max}}$ halos, while
the green lines correspond to low-$V_{{\rm max}}$ halos. 
The solid lines correspond to the full halo sample while the dashed line corresponds to the sample where ejected halos are removed. 
Lower panel: Ratio of linear biases between 
high-$V_{{\rm max}}$ and low-$V_{{\rm max}}$ samples from the Bolshoi
and the MultiDark simulations. The clustering difference increases
with decreasing $\mvir,$ and exceeds $30\%$ for halos of Milky Way
mass $\mvir\approx10^{12}\msun.$}
\end{figure}

The top panel of Fig.~\ref{fig:linear-bias} shows $\bhlin$ as a
function of $\mvir;$ results for high-$\vmax$ halos are shown in
blue, low-$\vmax$ halos in green. At the low-mass end, linear bias
is a weak function of $\mvir;$ for  $M_{{\rm vir}}\gtrsim\mcoll\approx10^{12.8}\msun,$ where $\mcoll$ is a characteristic mass scale for clustering corresponding to $\sigma(\mcoll, z) = \delta_{c} \approx 1.69$, we
see that $\bhlin$ increases sharply with $\mvir.$ Thus the basic
shape of each curve in Fig. \ref{fig:linear-bias} is in accord with
theoretical expectations from the peak-background split \citep{sheth_tormen99}
formalism and Press-Schechter theory \citep{pressschechter74}.
By comparing the blue and green curves in Fig.~\ref{fig:linear-bias}
we can see that linear bias has significant dependence upon $\vmax$
for halos of the same mass. This dependence is most readily seen in
the bottom panel, which shows the ratio of $\bhlin$ of high-$\vmax$
samples divided by $\bhlin$ of low-$\vmax$ samples. Thus in the
bottom panel of Fig.~\ref{fig:linear-bias}, vertical axis values
exceeding unity correspond to masses where high-$\vmax$ halos cluster
more strongly relative to their low-$\vmax$ counterparts.

At the low-mass end, high-$\vmax$ halos are more strongly clustered
than low-$\vmax$ halos of the same mass. The clustering difference
increases with decreasing $\mvir,$ and exceeds $30\%$ for halos
of Milky Way mass $\mvir\approx10^{12}\msun.$ At the high-mass end,
the trend reverses, and the overall magnitude is weaker. These results
are consistent with \citet{wechsler06}, who find that the same trends
hold when the secondary halo property is NFW concentration, rather
than $\vmax.$ This agreement is to be expected: insofar as the halo
profile is well-approximated by an NFW profile, $\vmax$ is entirely
determined by concentration (see Eq.~\ref{eq:vmax-mvir}). Note that there is a drop in the ratio of
the linear biases on the low-mass end. As we see no physical reason for this drop, we consider this to be 
a resolution effect and suggest that the completeness requirements for two-halo-property-dependent clustering 
are significantly more stringent relative to the requirements demanded by the need for a complete halo mass/velocity function.

Next, we study the scale-dependence of halo bias on small scales for high-$V_{{\rm max}}$
and low-$V_{{\rm max}}$ halos. We wish to parse the novel, small-scale effects from the 
well-known large-scale effects. So for each sample of high- and low-$V_{{\rm max}}$ halos, 
we compute the following quantity:

\begin{equation}
\tilde{b}_{\rm h}(r | \mvir; \vmax) \equiv b_{\rm h}(r | \mvir; \vmax) / b_{\rm lin}(\mvir; \vmax).
\end{equation}
Thus for any sample of halos, as $r\gtrsim10\mpch,$ we have $\tilde{b}_{\rm h}(r)\rightarrow1,$ by construction. 

\begin{figure}
\begin{center}
\includegraphics[width=0.9\columnwidth]{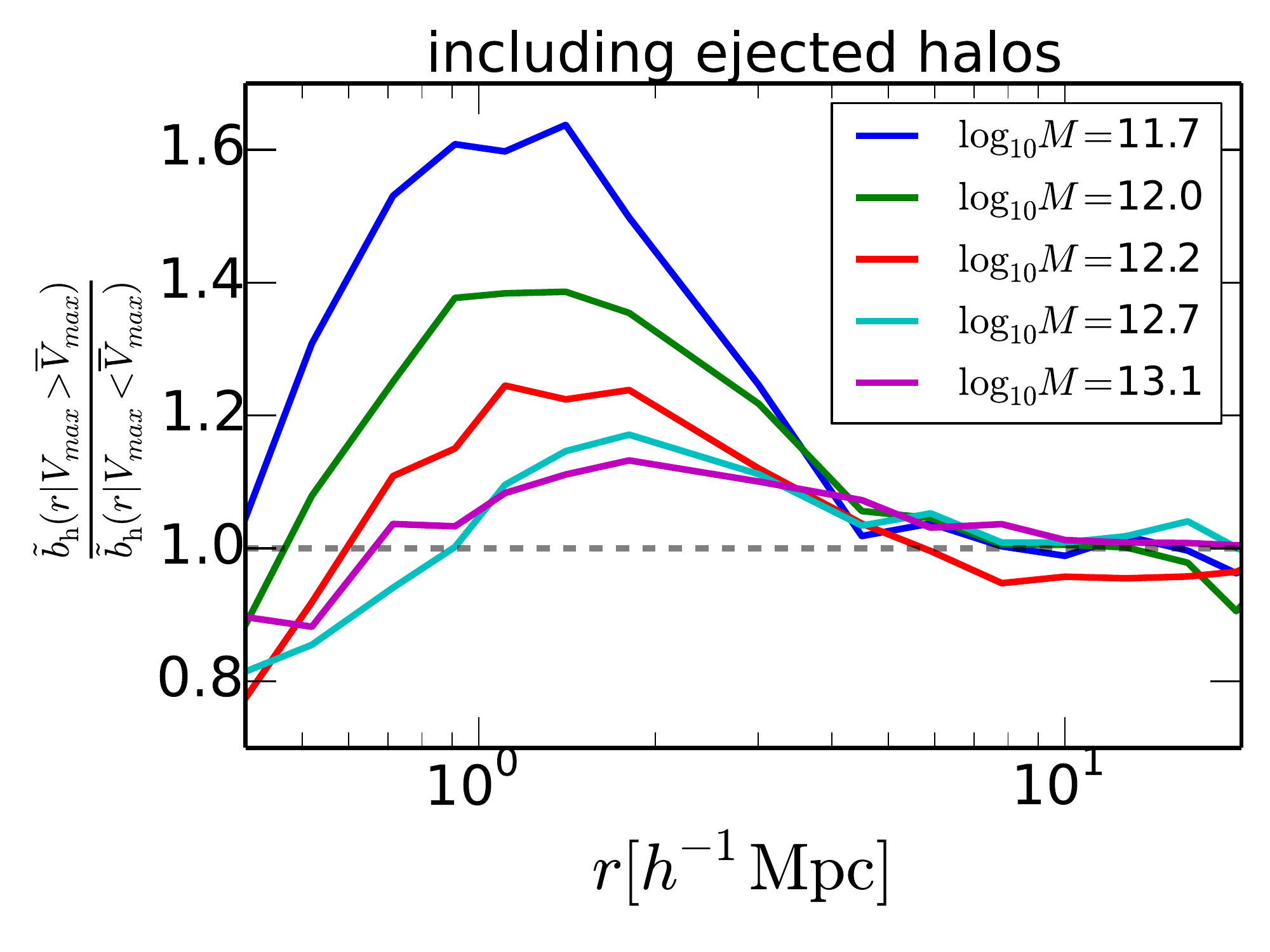}
\includegraphics[width=0.9\columnwidth]{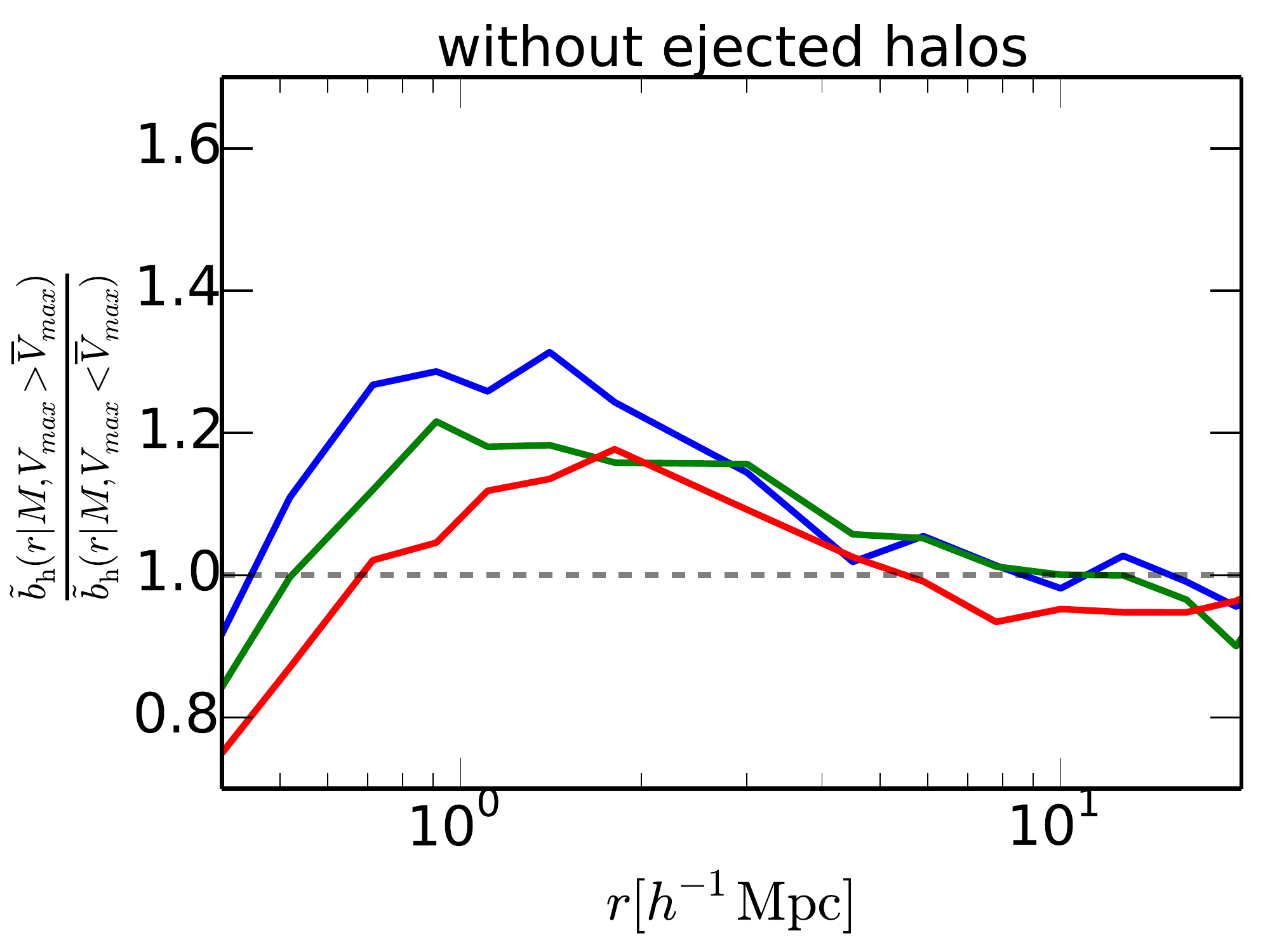}
\end{center}
\caption{\label{fig:small-scale} Top panel: Ratio of halo-matter cross correlation functions
between high/low-$V_{{\rm max}}$ halos from the Bolshoi simulation and the MultiDark
simulation at $z=0$, normalized by their linear
biases. Each line corresponds to different halo mass
bins labeled in the plots. The top three lines that correspond to
low mass halos are computed from halos in the Bolshoi simulation,
while the bottom lines are from the MultiDark simulation. Those plots
show that high-$V_{{\rm max}}$ halos cluster more strongly than low-$V_{{\rm max}}$
halos at $1h^{-1}{\rm Mpc}$ and the relative scale-dependence between
those subsamples increases smoothly with decreasing halo mass. Bottom panel: The same figure 
as the top panel without ejected halos only from the Bolshoi simulation. As can be seen by 
comparing these results to those in the top panel, the $V_{{\rm max}}$-dependence 
of halo bias on small scales is dramatically reduced by excluding ejected subhalos. This 
implies an intimate connection between the scale-dependence of assembly bias and subhalo back-splashing.}
\end{figure}

The bottom panel of Fig. \ref{fig:small-scale} shows
the ratio of $\tilde{b}_{h}(r)$ of high-$V_{{\rm max}}$ samples
divided by $b_{h}(r)$ of low-$V_{{\rm max}}$ samples for several
mass bins. The first three mass bins labeled in the figure, $M_{{\rm vir}}=10^{11.7,12.0,12.2}h^{-1}{\rm {\rm M_{\odot}}}$,
are from the Bolshoi simulation, and the last two mass bins, $M_{{\rm vir}}=10^{12.7,13.1}h^{-1}{\rm {\rm M_{\odot}}}$,
are from the MultiDark simulation. High-$V_{{\rm max}}$
halos cluster more strongly compared to their low-$V_{{\rm max}}$
counterparts at $1h^{-1}{\rm Mpc}$. This scale-dependent feature
becomes stronger with decreasing $M_{{\rm vir}}$ and exceeds 40\%
for halos of Milky Way mass and reaches 60\% at $M_{{\rm vir}}\approx10^{11.7}h^{-1}{\rm M_{\odot}}$.

Up until present, we have used both host halos and ejected halos to
compute halo biases. Both types of halos are identified as distinct
halos at $z=0$. Ejected halos, however, are halos which were identified
as part of more massive halos at one or more occasions in the past,
but were ejected and now exist as a host halo at $z=0$. Those ejected
halos tend to exist around more massive halos \citep[e.g.]{wetzel_etal13, wang09}. Therefore,
the effect on scale-dependent biases may be caused by those ejected
halos. 

To test this ejected halo hypothesis, we compute halo-matter cross correlation functions
after first excluding the subpopulation of ejected halos. 
Our results for the linear regime are shown as the dashed curves in 
Fig. \ref{fig:linear-bias}. The relative difference in the
linear bias between high-$V_{{\rm max}}$ and low-$V_{{\rm max}}$
halos is suppressed to 25\% for halos of Milky Way mass. This suppression due 
to excluding the ejected halos is consistent with the results presented in \cite{wang09}. 

In the bottom panel of Fig. \ref{fig:small-scale}, we show the scale-dependence of assembly bias for non-ejected halos.\footnote{We remind the reader that due to the mass resolution of the MultiDark simulation, fig. \ref{fig:small-scale} only shows results for Bolshoi halos.}
Once the ejected halos have been removed, the scale-dependent feature of assembly bias is greatly reduced. This implies an intimate
connection between the scale-dependence of assembly bias and subhalo back-splashing (see \S\ref{sec:discussion} for further discussion).

\section{Observational Consequences}

We now consider possible observational consequences of the results
of the previous section. In order to do this, the key first step is
to relate an observable property of a galaxy (luminosity, stellar
mass) to an intrinsic property of its host halo (mass, circular velocity).
As one might infer from above (and we demonstrate below), different
choices for the latter can result in significant differences for different
observables.

In order to be explicit, we use the abundance matching technique \citep{kravtsov04a,vale_ostriker04,tasitsiomi_etal04,vale_ostriker06,conroy_wechsler09,
guo10,simha10,neistein11a,watson_etal12b,rod_puebla12,kravtsov13} to connect the
stellar masses of central galaxies to either the mass or circular
velocity of host halos. We implement this by splitting the halo catalog
into a series of bins with constant number density (=$1.6\times10^{-3}(h^{-1}{\rm Mpc})^{-3}$),
rank ordering either by mass or circular velocity. We label each bin
by its corresponding stellar mass, computed from the stellar-to-halo
mass relation of \citep{behroozi13}. 

Note that when we rank order based on
circular velocity, there is the possibility that the mean halo
masses of these bins could differ from what we obtain after rank ordering
by halo mass. We explicitly check this and find that the mean
halo masses for both cases agree to $\sim99.6\%$, allowing us to consistently
compare samples of mock central galaxies with the same stellar mass, but where 
the stellar mass is statistically regulated by either $\mvir$ or $\vmax.$ 

We find only a relatively minor difference in the large-scale clustering of the two samples of mock central galaxies. 
At fixed stellar mass, the linear bias of samples selected by their
circular velocity are $\sim5\%$ higher than samples selected by halo
mass. This decreases to $\sim2\%$ if we remove ejected halos from
both samples. 

\begin{figure}
\begin{center}
\includegraphics[width=0.9\columnwidth]{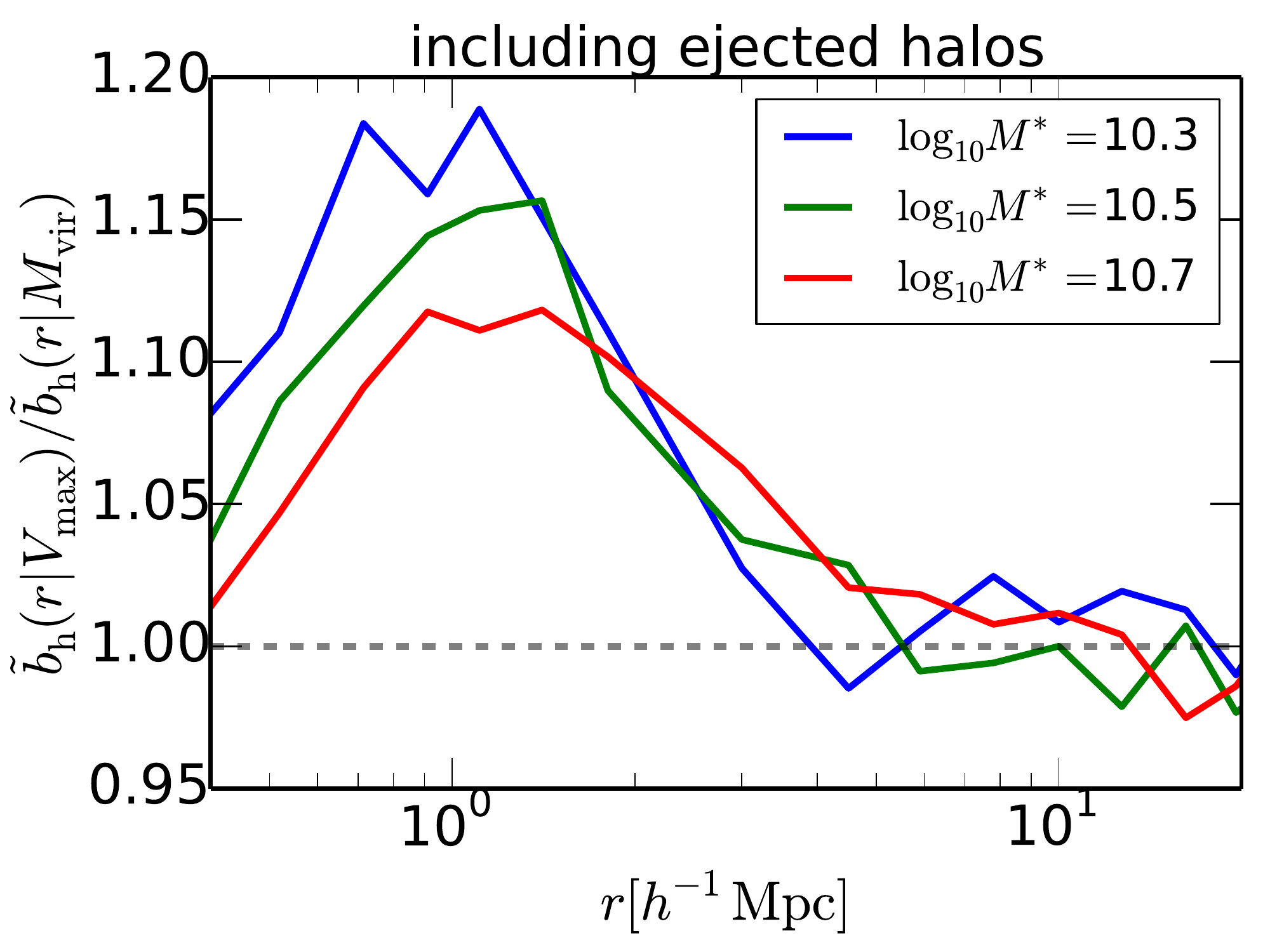}
\includegraphics[width=0.9\columnwidth]{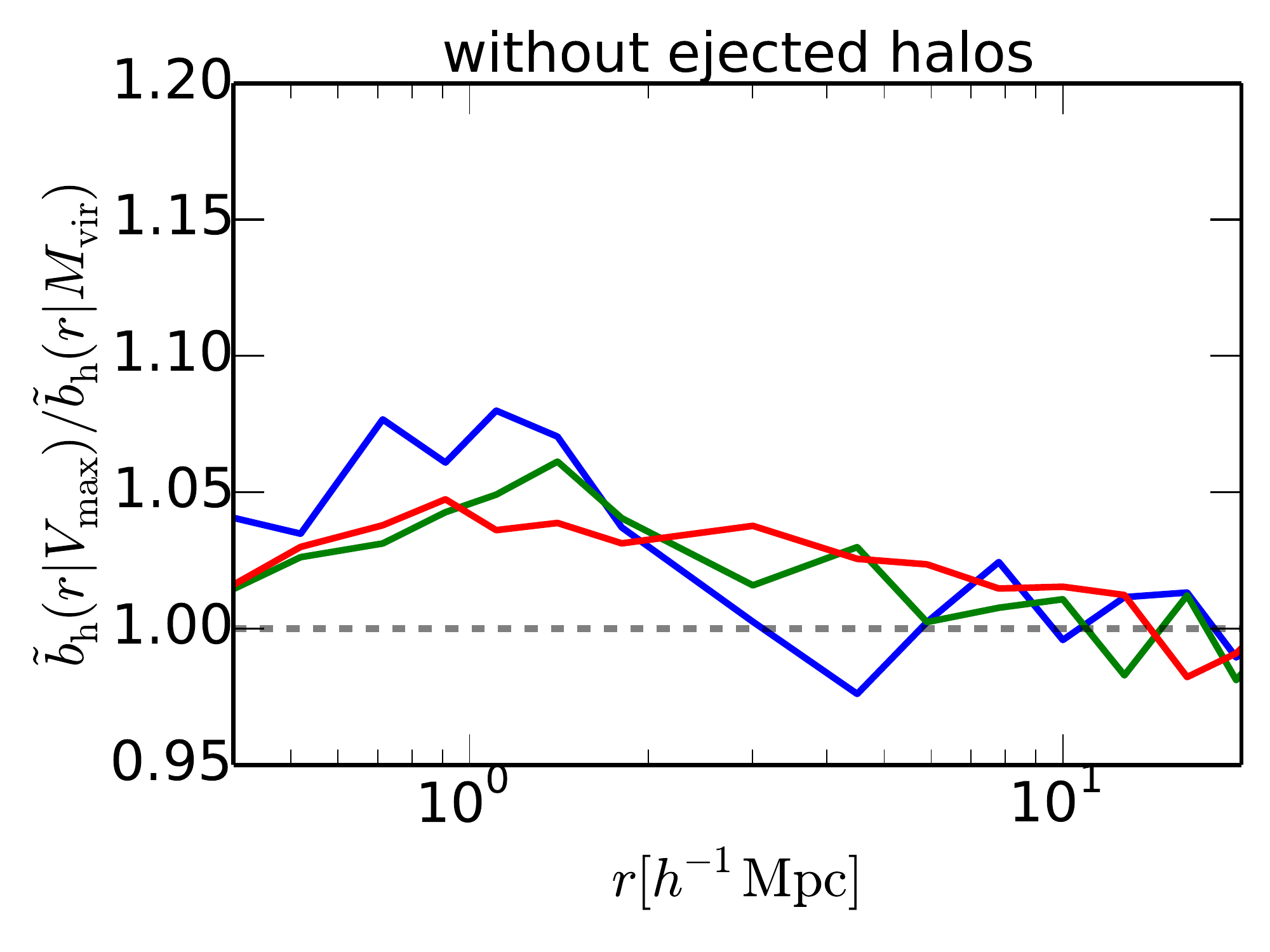}
\end{center}
\caption{\label{fig:abundance_small}Ratio of halo-matter cross correlation
functions between $M_{{\rm vir}}$-based and $V_{{\rm max}}$-based
abundance matching samples with ejected halos (top) and without ejected
halos (bottom). The ratios are normalized by their linear biases. Each
line corresponds to different stellar mass bins labeled in the plots.
With ejected halos, the halo biases computed from the $V_{{\rm max}}$-selected samples
have very different scale-dependence than the ones from the $M_{{\rm vir}}$-selected samples.
By removing those ejected halos, the difference is reduced by threefold.}
\end{figure}

We study the scale-dependence of the clustering of our mock central galaxies in Fig.~\ref{fig:abundance_small}, 
which is directly analogous to Fig. \ref{fig:small-scale}, only here we have use the abundance matching technique described above 
to illustrate how our ``halo-level" results may manifest in observed galaxy populations. 
Again we see a clear scale-dependence of
the clustering signal, with a maximum difference of $\sim15\%$ at $\sim1h^{-1}{\rm Mpc}$.
These differences go down to $\sim5\%$ after removing ejected subhalos, again reflecting the intimate connection 
between scale-dependent assembly bias and subhalo back-splashing.

\section{Discussion}
\label{sec:discussion}

For halos $M_{{\rm vir}}\gtrsim\mcoll\approx10^{12.8}\msun,$ we have shown that the linear bias of low-$V_{{\rm max}}$ halos 
is larger relative to high-$V_{{\rm max}}$ halos of the same mass. As shown in \citet{dalal_etal08}, this 
phenomenon is nicely explained in terms of the statistics of fluctuations in a Gaussian random field.
Consider two halos with the same present-day mass, but with different concentration. Both halos originate from a 
 fluctuation of the same peak height, but with different peak curvature: the high-concentration (high-$\vmax$) halo 
has a sharper peak than the low-concentration (low-$\vmax$) halo. 
\citet{dalal_etal08} showed that a generic prediction of Extended Press Schechter theory (EPS) with a configuration space filter 
is that low-curvature peaks cluster more strongly relative to high-curvature peaks of the same height. In closely related work, 
 \citet{zentner07} used EPS with a configuration space filter to show that for a pair of halos of the same peak height, the early-forming halo should 
 reside in a denser large-scale environment than the late-forming one.\footnote{See Chapter IX, Section D.}
 
A critical assumption underlying these EPS predictions is that a halo is the dominant peak in its large-scale environment. This is a well-founded assumption at the high-mass end, and we see that the predictions are in good agreement with simulations in the $\mvir>\mcoll$ regime \citep{dalal_etal08}. The situation is quite different when $\mvir\lesssim\mcoll.$ We have confirmed previous results \citep[e.g.,][]{wechsler06} that large-scale assembly bias changes sign and strengthens for lower-mass halos. This is in stark contrast to the EPS model described above, which makes the same prediction for assembly bias regardless of halo mass. 

Thus EPS succeeds and fails in precisely the regimes where we expect. Lower-mass halos are strongly influenced by the tidal field in which they evolve \citep{hahn_etal07b,wang_etal11,shi15,hahn_etal09,hearin_etal15}; the EPS assumption that the halo dominates its environment breaks down catastrophically, and in this regime nonlinear evolution governs assembly bias. On the other hand, high-mass halos {\em do} dominate their tidal environment; the EPS assumption holds good, and we can understand assembly bias as naturally arising from the statistics of Gaussian fluctuations. 

Our results on the scale-dependence of assembly bias are also consistent with this picture. 
First, we remind the reader that halo bias for high-$\vmax$ halos
shows non-trivial scale dependence with a pronounced bump at $\sim1-2\mpch$
compared to the bias for low-$\vmax$ halos. This scale-dependent
feature for high-$\vmax$ halos becomes 60\% larger compared
to low-$\vmax$ halos at $M_{{\rm vir}}\approx10^{11.7}\msun$. 
This feature, however, is removed by excluding the ejected halos, 
implying that this special sub-population is responsible for the scale-dependent bump. 

This scale-dependence has a simple interpretation in terms of subhalo back-splashing. First, ejected halos are
physically associated with the more massive halo from which they were ejected.
The clustering of ejected halos is therefore largely determined by this associated massive halo, much like the clustering
of present-day subhalos is determined by their host halo. Second, as ejected halos pass near and inside a massive halo, their physical growth is arrested, and many such halos even experience substantial mass loss \citep{wang09, wetzel_etal13}. This arrested development has a greater impact on the outer layers of the halo, 
so that the ejected halo's mass is significantly more affected than its circular velocity \citep{behroozi13c}. Putting these two effects together, we should naturally expect the outskirts of massive groups and clusters to be preferentially populated with low-mass halos that have above-average values of $\vmax$ for their mass. This manifests in the scale-dependent bump shown in Figures \ref{fig:small-scale} \& \ref{fig:abundance_small}. 


Recent advances in our understanding of halo growth sheds further light on the above results. As shown in \citet{diemer14,adhikari14,wetzel_nagai14,more15}, the natural physical boundary of a dark matter is the so-called ``splashback radius", which is the radius where accreted matter reaches its first apocenter after turnaround, and is roughly $2-3\rvir.$ As shown in \citet{wetzel_etal13}, the halos of massive groups and clusters ($\mvir\gtrsim10^{13}\msun,$ $\rvir\gtrsim500\kpch$) are surrounded by a large fraction of ejected halos. For ejected halos with $\mvir\approx10^{11.7}h^{-1}M_{\odot}$ halos, the host-centric spatial distribution of the ejected population peaks at $r\approx1.25R_{\rm 200m}\approx1.5\mpch$ (see Figs. 2 and 3 in \cite{wetzel_etal13}). The bump feature we find at $1-2\mpch$ is therefore in quantitative agreement with the \citet{wetzel_etal13, wang_mo_jing08} results: this bump occurs at the same physical scale that we would expect if the clustering of the ejected population is largely determined by the host to which the halos are ultimately bound. Note that the assembly bias on large scales, $r\gtrsim10\mpch,$ is not dominated by the ejected halos, which is consistent with \cite{wang_mo_jing08}. Without the ejected halos, the relative difference in the
linear bias between high-$V_{{\rm max}}$ and low-$V_{{\rm max}}$
halos remains 25\% for halos of Milky Way mass.

In order to explore possible observational consequences of our findings, we
use abundance matching relating a stellar mass of central galaxies
to either mass or circular velocity of host halos. Using different
intrinsic properties for host halos results in significant differences
in clustering signals of central galaxies. On large scales, the linear
bias of samples selected by their circular velocities are $\sim5\%$
higher than samples selected by halo masses at fixed stellar mass.
Without the ejected halos, this differences is reduced to $\sim2\%$.
On small scales ($r<10h^{-1}{\rm Mpc}$), samples selected by their
circular velocities exhibit the scale dependence with a bump at $1-2h^{-1}{\rm Mpc}$
compared to samples selected by halo masses. This scale-dependent
bump for $V_{{\rm max}}-$selected samples becomes $\sim15\%$ with
the ejected halos and $\sim5\%$ without the ejected halos. As these effects are 
roughly as large as existing SDSS clustering measurements on these scales, 
this raises the possibility that clustering measurements can be used to determine which host halo 
property is the true statistical regulator of the stellar mass of the central galaxy residing in the halo. 

\section{Summary}
\label{section:summary}

We conclude the paper with an overview of our primary results:

\begin{enumerate}
\item At fixed mass $\mvir,$ the large-scale bias of halos exhibits significant residual dependence on potential well depth $\vmax.$ At the low-mass end, high-$\vmax$ halos cluster more strongly than their low-$\vmax$ counterparts. At the high-mass end, this trend reverses, and is generally weaker, with the transition occurring at $\mcoll\approx10^{12.5}\msun.$ Our results are quantitatively consistent with previous studies of large-scale halo assembly bias. 
\item We show that assembly bias exhibits complex scale-dependence. The $\vmax-$dependence of halo clustering shows a pronounced ``bump" on scales $500\kpch\lesssim r\lesssim 5\mpch.$ This scale-dependence is itself mass-dependent: the bump feature is strongest for low-mass halos and vanishes for halos with $\mvir\gtrsim\mcoll.$
\item The scale-dependence of assembly bias can primarily be attributed to a special sub-population of {\em ejected subhalos}, which experience arrested mass-growth before and after being ejected from a higher-mass host. If this special population is excluded from the halo sample, the strength of small-scale assembly bias is limited to $\lesssim5\%$ for all masses $\mvir\gtrsim10^{11.75}\msun.$
\end{enumerate}

\section{Acknowledgements}

We thank Andrew Zentner and Frank van den Bosch for informative conversations. A portion of this work was supported by the National Science Foundation under grant PHYS-1066293 and the hospitality of the Aspen Center for Physics.

\bibliographystyle{mn2elong}
\bibliography{VmaxMvir}

\end{document}